\documentclass[12pt]{article}
\usepackage{epsfig}
\usepackage{amssymb}
\usepackage{pslatex}

\makeatletter

\@addtoreset{equation}{section}
\makeatother

\newcommand{\bea}{\begin{eqnarray}}
\newcommand{\eea}{\end{eqnarray}}
\newcommand{\be}{\begin{eqnarray}}
\newcommand{\ee}{\end{eqnarray}}
\newcommand{\nn}{\nonumber}

\def\fR{{\mathfrak R}}

\def\pa{{\phi^a}}
\def\pb{{\phi^b}}
\def\ta{{{\tilde{\phi}}^\alpha}}
\def\tb{{{\tilde{\phi}}^\beta}}
\def\pad{{\phi^\dagger_a}}
\def\pbd{{\phi^\dagger_b}}
\def\tad{{{\tilde{\phi}}^\dagger_\alpha}}
\def\tbd{{{\tilde{\phi}}^\dagger_\beta}}
\def\sa{{\psi^\alpha}}
\def\sb{{\psi^\beta}}
\def\ua{{{\tilde{\psi}}^a}}
\def\ub{{{\tilde{\psi}}^b}}
\def\sad{{\psi^\dagger_\alpha}}
\def\sbd{{\psi^\dagger_\beta}}
\def\uad{{{\tilde{\psi}}^\dagger_a}}
\def\ubd{{{\tilde{\psi}}^\dagger_b}}
\def\fR{{\mathfrak R}}
\def\fL{{\mathfrak L}}
\def\fQ{{\mathfrak Q}}
\def\fS{{\mathfrak S}}

\begin{document}

\begin{titlepage}
\begin{flushright}
%SNUST 091102\\
UOSTP 120301
\end{flushright}
\vskip0.2cm
\centerline{\Large
\bf  Exact 4-point Scattering Amplitude}
\centerline{\Large\bf
of} \centerline{\Large\bf \ \ \ \ \
the Superconformal
 Schr\"odinger Chern-Simons Theory}

\vskip1.5cm
\centerline{Sangnam Park and
Dongsu Bak 
}
\vspace{1.0cm}
\centerline{\sl %$^b$
 Department
of Physics, University of Seoul, Seoul 130-743 Korea}
\centerline{( \tt dsbak@uos.ac.kr )}
\vspace{1.5cm}
\centerline{ABSTRACT} \vspace{0.75cm} \noindent

\noindent
We consider the non-relativistic superconformal $ U(N)\times U(N)$ Chern-Simons theory with level $(k,-k)$
possessing fourteen supersymmetries.
We obtain an exact four-point scattering amplitude of the theory to all orders in $1/N$ and
$1/k$ and prove that the scattering amplitude becomes trivial when $k=1$ and $2$. We confirm this amplitude
to one-loop order by using an explicit field theoretic computation and show that the beta function for the
contact interaction
vanishes to the one-loop order, which is consistent with the quantum conformal invariance
of the underlying theory.

\vskip1.25cm

\end{titlepage}

\section{Introduction}

Recently there has been much interest in the non-relativistic version of the AdS/CFT correspondence,
which has a potential application in some condensed matter systems in the strongly coupled region \cite{Son:2008ye}.
Some candidates for such non-relativistic conformal field theories can be
obtained by taking appropriate non-relativistic limits of the mass-deformed
 Aharony-Bergman-Jafferis-Maldacena (ABJM) theory.
The ABJM theory is the three dimensional
${\cal N}=6$ $U(N)\times U(N)$ superconformal Chern-Simons theory with level $(k,-k)$ and dual to
the type IIA string theory on the $AdS_4\times \mathbb{CP}_3$ background \cite{Aharony:2008ug}. Some tests of
this duality have
been carried out largely based on the integrability technique \cite{Minahan:2008hf}.

In this note, we shall study the non-relativistic version of the ABJM theory with fourteen
supersymmetries \cite{Nakayama:2009cz,Lee:2009mm}. This follows from the mass-deformed ABJM theory \cite{Hosomichi:2008jb,Gomis:2008vc} by taking the so-called `PPPP'
non-relativistic  limit where one chooses all Schr\"odinger fields from a particle sector instead of
an anti-particle sector \cite{Nakayama:2009cz,Lee:2009mm}. The resulting theory possesses $\mathfrak{psu}(2|2)$ symmetry (involving
eight kinematical supercharges), together with two extra kinematical and four conformal
supersymmetries, leading to fourteen supersymmetries in total. There is also the so-called `PAAP'
non-relativistic limit \cite{Nakayama:2009cz} where one chooses half of the  Schr\"odinger fields from the anti-particle
sector in an appropriate
manner. This theory has  only eight kinematical supersymmetries of the $\mathfrak{psu}(2|2)$.
There are non-relativistic limits leading to even less supersymmetric theories \cite{Nakayama:2009cz}.
For the other related aspects
of non-relativistic supersymmetric Chern-Simons theories, see Refs. \cite{okab,Colgain:2009wm,Kim:2009ny}.

In this note, we focus on the non-relativistic superconformal $U(N) \times U(N)$ Chern-Simons
theory with  level $(k,-k)$, which has fourteen supersymmetries, and compute its four-point scattering
amplitudes describing $2\rightarrow 2$ scattering processes. Adopting the previously developed method \cite{nacs},
we derive the two-body Schr\"odinger equation by which any $2\rightarrow 2$ scattering
processes can be described. Starting from its scattering solution, we shall obtain an exact four-point scattering
amplitude of any combination of incoming two-particle states to all orders in $1/N$ and $1/k$.
Because we are dealing with a Chern-Simons gauge theory, there is the so-called statistics interaction,
which corresponds
to the Aharonov-Bohm  %two-body
interaction  (between anyonic particles) characterized by
a phase $e^{i \pi \Omega}$, when two particles are exchanged. (The statistics interaction matrix
$\Omega$ will be specified below.) In addition, there is a two-body contact interaction between
particles that can be fully specified by a contact interaction matrix $C$. These two interaction
matrices, in fact, encode the complete interaction structure  of our non-relativistic Chern-Simons theory.
Any scattering amplitudes can be represented as functions of these two matrices.

We check this four-point amplitude perturbatively by using a direct field theoretic computation to one-loop
order. We shall show that the beta function for the contact interaction vanishes
to the one-loop order, which is consistent with the quantum conformal symmetry of
the underlying theory.  After %taking care of the
renormalization, the resulting amplitude
 to the one-loop order agrees precisely with %the prediction of
 our exact amplitude.

 In Section 2, we introduce the non-relativistic ABJM theory with fourteen supersymmetries
 and discuss the detailed structure of  $\mathfrak{psu}(2|2)$ symmetry. Especially, we write
 the contact  and the statistics interaction matrices in  manifestly
 $\mathfrak{psu}(2|2)$ covariant forms. In Section 3, we shall derive the two-body Schr\"odinger
 equation that describes generic $2\rightarrow 2$ scattering processes.
Section 4 deals with the scattering solution of the two-body Schr\"odinger
 equation. Using this solution, we shall extract the exact four-point scattering amplitude. We
 shall show that the scattering amplitude becomes completely trivial when $k=1$ and $2$. In Section 5,
 we perform a field theoretic perturbative analysis to the one-loop order.  We shall show that the beta
 function for the contact interaction vanishes for our form of the contact interaction as
 dictated by the
 quantum conformal invariance.
 We shall also show that the perturbative result agrees precisely with %that of
 the exact amplitude.
 The last section is devoted to concluding remarks.
 Some details of the non-relativistic ABJM theory with fourteen supersymmetries are presented in the Appendix.

\section{$\mathfrak{psu}(2|2)$ invariance}

We shall begin with the non-relativistic Chern-Simons
Lagrangian given by
\bea
{\cal L} &=&   \, {k \over 4\pi}CS(A)-{k \over 4\pi}
CS(\overline{A})%\nonumber \\
+{\mbox{Tr}} %\big[
 \,\,\,
\Phi_A^{\dagger} \Big(\,i\,D_0+ {1\over 2m} \vec{D}^2\Big)\, \Phi_A
\, %\big]
%+
%{\Phi}^\dagger_{\tilde{A}}
%\Big(\,i\,D_0+ {1\over 2m} \vec{D}^2\Big) {\Phi}_{\tilde{A}}
%\, \Big]
\nonumber\\
&-& {1\over 4}\,\,(\Phi_{{\cal A}_1})^\dagger
(\Phi_{{\cal A}_2})^\dagger \,\,
C^{\,\,{\cal A}_1 {\cal A}_2}_{\,\,\, {\cal B}_1 {\cal B}_2}
\,\,\, \Phi_{{\cal B}_1}\,\,
\Phi_{{\cal B}_2}\,,
\label{lagrange}
\eea
whose detailed component form is presented in Appendix A.
Apart from the two gauge fields, $A_\mu$ and $\overline{A}_\mu$,
there are 8 kinds of complex matter
fields, $\Phi_I= (\phi^a\,|\, \psi^\alpha )$
and $\Phi_{\tilde{I}}= (\tilde\psi^a\,|\, \tilde\phi^\alpha )$ where
one has 4 bosons $\phi^a,\, \tilde{\phi}^\alpha$
and 4 fermions $\psi^\alpha,\, \tilde{\psi}^a$.
We  use indices $I,\,\,\,J$ for $(\phi^a\,|\, {\psi}^\alpha)$,
%indices
$\tilde{I}, \,\,\,\tilde{J}$ for
$(\tilde{\psi}^a\,|\,  \tilde{\phi}^\alpha)$ and
%indices
$A,\,\,\, B$ for the total eight flavors.
The calligraphic upper case letters run %ning
over
this 8 flavored matrix component space whose total dimension
is $8 N^2$. In other words,
the index ${\cal A}$ represents $(Amn)$; {\it i.e.},
$\Phi_{\cal A}=\Phi_{A\,\,mn}\,$.
The contact interaction matrix $C$ acts upon  the two-body state
$|{\cal B}_1 {\cal B}_2\rangle$ in the
%component $|{\cal B}_1 {\cal B}_2\rangle$ in the
space of $\Phi_1\otimes \Phi_2$, which ends up with a new
two-body state $|{\cal A}_1 {\cal A}_2\rangle$, which is a
$64N^2\times 64 N^2$ matrix, whose detailed form will be specified
below.   This contact interaction matrix
will serve as a basic building block of our 4-point %$2\rightarrow 2$
scattering
amplitude.

Before specifying a detailed form of $C$,
let us first state the $\mathfrak{psu}(2|2)$ invariance of the above
Lagrangian.
The off-shell $\mathfrak{psu}(2|2)$
superalgebra  %of the excitation symmetry
is spanned by the two su(2) rotation generators
${\fR}^a\!_b$, ${\fL}^\alpha\!_\beta$,
the supersymmetry generator ${\fQ}^\alpha_a$ and
the superconformal generator ${\fS}^a_\alpha$.
The off-shell configuration is
characterized by $sl(2, \mathbb{R})$
central charges
$\mathfrak{C}, \mathfrak{K}, \mathfrak{K}^*$. %~\cite{Beisert:2005tm}.
Their %(anti)
commutators are  given by %\cite{Beisert:2005tm}
\bea
&& [{\fR}^a\!_b, \,\, \mathfrak{J}^c]= \delta^c_b \,
\mathfrak{J}^a-{1\over 2}
\delta^a_b \, \mathfrak{J}^c\,,\ \ \
[{\fL}^\alpha\!_\beta, \,\, \mathfrak{J}^\gamma]= \delta^\gamma_\beta  \,
\mathfrak{J}^\alpha-{1\over 2}
\delta^\alpha_\beta \, \mathfrak{J}^\gamma\,,
\nonumber\\
&& \{ {\fQ}^\alpha_a, \,\, \fS^b_\beta\}= \delta^b_a   {\fL}^\alpha\!_\beta
+ \delta^\alpha_\beta \, {\fR}^b\!_a
+ \delta^b_a  \delta^\alpha_\beta \mathfrak{C}\,,            \nonumber \\
&& \{ {\fQ}^\alpha_a, \,\, \fQ^\beta_b\}=\epsilon^{\alpha\beta}
\epsilon_{ab} \mathfrak{K}\,,\ \ \
\{ {\fS}^a_\alpha, \,\, \fS_\beta^b\}=\epsilon_{\alpha\beta}
\epsilon^{ab} \mathfrak{K}^* \,.
\eea
%The central charges $\mathfrak{C}$ is related to the energy
%by $E= \mathfrak{C}$, while $\mathfrak{K},
%\mathfrak{K}^*$ introduced at off-shell are related to the momentum.
On a state of fundamental
representations, the generators act as
\bea
&& \fR^a\!_b |\phi^c \rangle =
\delta^c_b  |\phi^a \rangle
-{1\over 2} \delta^a_b  |\phi^c\rangle
\,, \ \ \
\fL^\alpha\!_\beta |\phi^\gamma \rangle =
\delta^\gamma_\beta  |\phi^\alpha \rangle
-{1\over 2} \delta^\alpha_\beta  |\phi^\gamma\rangle
\eea
and as
\bea
&&\fQ^\alpha_a |\phi^b \rangle \, = \,
a \, \delta_a^b  |\psi^\alpha \rangle\,, \nonumber \\
&&
\fQ^\alpha_a |\psi^\beta \rangle =\,
b \, \epsilon^{\alpha\beta}\epsilon_{ab}  |\phi^b %G(A^+)
\rangle\,,\nonumber\\
&&\fS^a_\alpha|\phi^b \rangle \, =\,
c \, \epsilon_{\alpha\beta}\epsilon^{ab} |\psi^\beta %{G}({A^-})
\rangle \,, \nonumber \\
&& \fS^a_\alpha |\psi^\beta \rangle =\,
d \, \delta^\beta_{\alpha}  |\phi^a \rangle\,.
\label{repre}
\eea
Closure of the superalgebra on a fundamental representation
leads to the shortening condition
\bea
a d-b c=1 \, . \label{shortening1}
\eea
From the algebra,
one finds that $\mathfrak{K}=ab$, $\mathfrak{K}^*=cd$
and
$\mathfrak{C}={1 \over 2} (ad + bc)$. Unitarity of the representation
requires that $(\fQ^\alpha_a)^\dagger= \fS_\alpha^a$ and
$ab=(cd)^*$.

We have one more set of the fundamental representation where
the set $\Phi_I=(\phi^a \,| \,\psi^\alpha)$ in the above is replaced by
$\Phi_{\tilde{I}}=(\tilde{\psi}^a\,|\,\tilde{\phi}^\alpha)$
with a further replacement of $(a,b,c,d)$ by
$(\tilde{a},\tilde{b},\tilde{c},\tilde{d})$.
For  $\Phi_I$, we have $a=d=1$ and $b=c=0$ while
 $\tilde{a}=\tilde{d}=-1$ and $\tilde{b}=\tilde{c}=0$
for $\Phi_{\tilde{I}}$.

To show the $\mathfrak{psu}(2|2)$ invariance of the Lagrangian,
let us begin with the contact interaction part. For the specification of
the contact interaction terms, we rearrange  the quartic potential
term in Eq.~(\ref{apotential}) to the normal ordered form
$\Phi^\dagger\,\Phi^\dagger\,\,\Phi\,\Phi$ and use the equal time
(anti-) commutation relations
\be
[\Phi_{\cal A}({\bf r},t)\,, \Phi_{\cal B}({\bf r}',t)]_\pm
\equiv \Phi_{\cal A}({\bf r},t)\,\Phi_{\cal B}({\bf r}',t)
-(-)^{F_{\cal A}\,F_{\cal B}}\,\,\Phi_{\cal B}({\bf r}',t)\,
\Phi_{\cal A}({\bf r},t)=0\,,
\ee
where $F_{\cal A}$ denotes the fermion number of the
field $\Phi_{\cal A}$.

The two-body interaction involving only $\Phi_I$ fields is govern by
the matrix
\be
S= \mathbb{S}\,\otimes\, ({\rm P}-\overline{\rm P})\,,
\ee
which has  only nonvanishing matrix components in
$\langle I_1 m_1n_1,\, I_2 m_2n_2 |\,S\, |
J_1 p_1q_1,\, J_2 p_2q_2
\rangle$.
The permutations
$P$ and $\overline{P}$ are,  respectively,
defined by
\bea
{\rm P}^{\,\,m_1n_1 m_2n_2}_{\,\,p_1q_1 p_2 q_2}=
\delta^{m_2}_{p_1} \delta^{m_1}_{p_2}
\delta^{n_1}_{q_1}\delta^{n_2}_{q_2}\,,\ \ \ \ \  \
{\overline{\rm P}}^{\,\,m_1n_1 m_2n_2}_{\,\,p_1q_1 p_2 q_2}=
\delta^{m_1}_{p_1} \delta^{m_2}_{p_2}
\delta^{n_2}_{q_1}\delta^{n_1}_{q_2}\,.
\eea
The matrix $\mathbb{S}$, which has  nonvanishing components only in
$\mathbb{S}^{I_1I_2}_{J_1J_2}$, can be identified as
\be
\mathbb{S}^{I_1I_2}_{J_1J_2}=
-{2\pi\over m\,k}\Big(\,
\mathbb{I}^{I_1I_2}_{J_1J_2}-\mathbb{P}^{I_1I_2}_{J_1J_2}
\,\Big)\,,
\ee
where the identity $\mathbb{I}$ and the graded permutation
$\mathbb{P}$ are, respectively, defined by
\be
\mathbb{I}^{\,B_1B_2}_{\,A_1A_2}=\delta^{B_1}_{A_1}
\delta^{B_2}_{A_2}\,, \ \ \ \ \ \ \
\mathbb{P}^{\,B_1B_2}_{\,A_1A_2}=(-)^{F_{A_1}\, F_{A_2}}\delta^{B_2}_{A_1}
\delta^{B_1}_{A_2}\,.
\ee
The graded permutation is $\mathfrak{psu}(2|2)$ invariant.
Therefore, $\mathbb{S}$ or $S$ are  $\mathfrak{psu}(2|2)$ invariant.
Similarly, for the contact interaction involving only
$\Phi_{\tilde{I}}$ fields, one has an interaction of the form
\be
\widetilde{S}=\widetilde{\mathbb{S}}\,\otimes\, ({\rm P}-\overline{\rm
P})\,,
\ee
where $\widetilde{\mathbb{S}}$ has
only the nonvanishing components
\be
\widetilde{\mathbb{S}}^{\tilde{I}_1\tilde{I}_2}_{\tilde{J}_1\tilde{J}_2}=
+{2\pi\over m\,k}\Big(\,
\mathbb{I}^{\tilde{I}_1\tilde{I}_2}_{\tilde{J}_1\tilde{J}_2}
-\mathbb{P}^{\tilde{I}_1\tilde{I}_2}_{\tilde{J}_1\tilde{J}_2}
\,\Big)\,.
\ee

For the interaction involving  $\Phi_I$ and $\Phi_{\tilde{I}}$
at the same time,
we introduce $\mathbb{T}$ defined by
\bea
&&{m\,k\over 2\pi}\, \mathbb{T}\,\,|\phi^a\tilde{\phi}^\beta\rangle
=-%\big( \,
|\tilde{\psi}^a{\psi}^\beta\rangle
-|{\psi}^\beta\tilde{\psi}^a\rangle
%\,
%\big)
\,,\ \ \ \
{m\,k\over 2\pi}\,
\mathbb{T}\,\,\,\,|\tilde{\phi}^\beta\phi^a\rangle
=|{\psi}^\beta\tilde{\psi}^a\rangle+
|\tilde{\psi}^a{\psi}^\beta\rangle\,,
\nonumber\\
&&{m\,k\over 2\pi}\,  \mathbb{T}\,\,|\psi^\alpha\tilde{\psi}^b\rangle
=
|\tilde{\phi}^\alpha{\phi}^b\rangle
-|{\phi}^b\tilde{\phi}^\alpha\rangle
\,,\ \ \ \  \  \ \
{m\,k\over 2\pi}\, \mathbb{T}\,\,|\tilde{\psi}^b\psi^\alpha\rangle
=
|\tilde{\phi}^\alpha{\phi}^b\rangle
-|{\phi}^b\tilde{\phi}^\alpha\rangle\,,
\nonumber\\
&& {m\,k\over 2\pi}\, \mathbb{T}\,\,|\phi^a\tilde{\psi}^b\rangle
=-
|\tilde{\psi}^a{\phi}^b\rangle
+|{\phi}^b\tilde{\psi}^a\rangle
\,,\ \ \ \  \  \,\,
{m\,k\over 2\pi}\, \mathbb{T}\,\,|\tilde{\psi}^b\phi^a\rangle
=-|{\phi}^b\tilde{\psi}^a\rangle+
|\tilde{\psi}^a{\phi}^b\rangle\,,
%\nonumber\\
\nonumber\\
&& {m\,k\over 2\pi}\,
\mathbb{T}\,\,|\tilde{\phi}^\alpha
\psi^\beta\rangle
=
|{\psi}^\alpha\tilde{\phi}^\beta\rangle
-|\tilde{\phi}^\beta {\psi}^\alpha\rangle
\,,\ \ \ \  \ \ \, \,
{m\,k\over 2\pi}\, \mathbb{T}\,\,|
\psi^\beta
\tilde{\phi}^\alpha\rangle
=|\tilde{\phi}^\beta {\psi}^\alpha\rangle-
|
{\psi}^\alpha
\tilde{\phi}^\beta\rangle\,.
\eea
This matrix is also $\mathfrak{psu}(2|2)$ invariant.
In summary,
the contact interaction matrix is  given by
\be
C=(\,\mathbb{S}+\widetilde{\mathbb{S}}+\mathbb{T}\,)\,\otimes\,
(\,{\rm P}-\overline{\rm P}\,)\,.
\ee
An interesting property we shall use later on is
\be
(\,\mathbb{S}+\widetilde{\mathbb{S}}+\mathbb{T}\,)^2 =
\Big({4\pi\over m\,k}\Big)^2
\,\,{
\mathbb{I}-\mathbb{P}
\over 2}\,.
\label{pro}
\ee
This construction ensures the
$\mathfrak{psu}(2|2)$ invariance of the contact interaction term in
Eq.~(\ref{lagrange}).

With $\delta A_\mu=0$ under the
$\mathfrak{psu}(2|2)$ transformation, the kinetic terms are also invariant
under the $\mathfrak{psu}(2|2)$ transformation. Thus, we conclude that
the system in Eq.~(\ref{lagrange}) has
$\mathfrak{psu}(2|2)$ invariance.
Finally, the tree-level gauge interaction is characterised by the
so-called statistical interaction matrix given by
\be
\Omega = {1\over k}\,\, \mathbb{I}\,\otimes
(\,
{\rm P}-\overline{\rm P}
\,)\,.
\ee
We note that
its strength is governed by the inverse of the Chern-Simons
level
and that flavors are not changed by this gauge interaction.

\section{Two-body Schr\"odinger equation}

For the Schr\"odinger Chern-Simons system, the Gauss law
constraints
\be
F_{12}(A)=\rho_A  \ \ \  {\rm and}\ \ \
F_{12}(\overline{A})=\rho_{\overline{A}}\,
\ee
can be solved explicitly in the gauge
$A_1=\overline{A}_1=0\,\,$ \cite{nacs}. One may then
eliminate the gauge fields by the solution,
\be
A_2={1\over \partial_1}\,\,\rho_A  \ \ \  {\rm and}\ \ \
\overline{A}_2={1\over \partial_1}\,\,\rho_{\overline{A}}\,,
\label{gsol}
\ee
by which one gets
 an equivalent system that only depends on the matter
fields. One may show that no nontrivial functional Jacobian
factor arises in this procedure \cite{nacs}. The resulting
action is simply given by
\bea
{\cal L} = {\mbox{Tr}} %\big[
 \,\,\,
\Phi_A^{\dagger} \Big(\,i\,\partial_0+ {1\over 2m} \vec{D}^2\Big)\, \Phi_A
\, %\big]
%+
%{\Phi}^\dagger_{\tilde{A}}
%\Big(\,i\,D_0+ {1\over 2m} \vec{D}^2\Big) {\Phi}_{\tilde{A}}
%\, \Big]
%\nonumber\\
- {1\over 4}\,\,(\Phi_{{\cal A}_1})^\dagger
(\Phi_{{\cal A}_2})^\dagger \,\,
C^{\,\,{\cal A}_1 {\cal A}_2}_{\,\,\, {\cal B}_1 {\cal B}_2}
\,\,\, \Phi_{{\cal B}_1}\,\,
\Phi_{{\cal B}_2}\,,
%\label{lagrange}
\eea
where the covariant derivative is defined with the solution
in Eq.~(\ref{gsol}).

The $n$-body Schr\"odinger equation can be derived using the
operator Schr\"odinger equation following a
straightforward procedure \cite{nacs}. The $2\rightarrow 2$ scattering
amplitude of our interest can be obtained using
this  two-body Schr\"odinger equation.
The two-body wave function is defined by
\be
\Psi_{{\cal A}_1 {\cal A}_2}({\bf r}_1,{\bf r}_2; t)
=\langle 0|
\Phi_{{\cal A}_1}({\bf r}_1,t)
\Phi_{{\cal A}_2}({\bf r}_2,t) |\Psi\rangle
=
\langle 0 |
\Phi_{ A_1\,m_1 n_1}({\bf r}_1,t)
\Phi_{A_2\, m_2 n_2}({\bf r}_2,t) | \Psi \rangle\,,
\ee
%This two-body wave function
which has the exchange symmetry
\be
\Psi_{{\cal A}_1 {\cal A}_2}({\bf r}_1,{\bf r}_2; t)
= (-)^{F_{{\cal A}_1} F_{{\cal A}_2}}
\Psi_{{\cal A}_2 {\cal A}_1}({\bf r}_2,{\bf r}_1; t)
=(\mathbb{P}\,{\rm P}\,\overline{\rm P}\,\,
\Psi)_{{\cal A}_1 {\cal A}_2}({\bf r}_2,{\bf r}_1; t)\,.
\ee

We find that the two-body Schr\"odinger equation takes
the following form:
\be
i\,\partial_t\,
{\Psi} ({\bf r}_1,{\bf r}_2; t)
=\Big[\!-\!{1\over 2m}\big(
\nabla_1 \!+\! 2\pi i\Omega\, {\bf G}({\bf r}_{12} %\!-\!{\bf r}_2
)\,
\big)^2 +\big(1\leftrightarrow 2\big)+{C\over 2}\,\delta
({\bf r}_{12}%\!-\!{\bf r}_2
)\,
\Big] \Psi ({\bf r}_1,{\bf r}_2; t)\,,
\ee
where $G_i({\bf r})= {1\over 2\pi}\epsilon_{ij}\partial_j \ln r$ and ${\bf r}_{12} ={\bf r}_1\!-\!{\bf r}_2$.
One thing to note is that  we have performed a non-singular
gauge transformation from the axial gauge to the Coulomb gauge
$\nabla\cdot {\bf G}=0$. In the center-of-momentum frame, the
time-independent Schr\"odinger equation becomes
\be
\Big[-{1\over m}\big(
\nabla + 2\pi i\,\Omega\, {\bf G}({\bf r})\,
\big)^2 +{C\over 2}\,\delta
({\bf r})-E\,
\Big] \Psi ({\bf r})=0\,,
\label{sch}
\ee
where, with relative momentum ${\bf p}$,
 we take $E= {p^2\over m}$ for our scattering problem.
This equation will be the
starting point of our analysis in the next section
for the construction of our scattering amplitude.

\section{Exact $2\rightarrow 2$ scattering amplitude}

Before solving our main problem for the full scattering
amplitude, let us first consider the single-component
Schr\"odinger   equation given by
\be
\Big[-{1\over m}\big(
\nabla + 2\pi i\,\nu\, {\bf G}({\bf r})\,
\big)^2 +{c\over 2}\,\delta
({\bf r})-{p^2\over m}\,
\Big] \psi ({\bf r})=0\,,
\ee
where $\nu$ and $c$ are not matrix-valued, but $c$-numbers.
This may be viewed as an
 eigen-component equation in the basis where
 $\Omega$ and $C$ are simultaneously
diagonalized.

The scattering solution with the boundary condition $\psi(0)=0$
reads \cite{aha,rui,jackiw,oren}
\be
\psi(r,\theta)= e^{ipr\cos \theta-i\, \nu(\bar{\theta}-\pi)}
-\sin \nu \pi\,\,
e^{-i ([\nu]+1)\theta}\, \int^\infty_{-\infty}
{dt\over \pi} e^{ipr \cosh t}
{e^{-\{\nu\}t}
\over
e^{-i \theta}- e^{-t}}\,,
\label{ssol}
\ee
where $[\nu]$ denotes the greatest integer part of $\nu$,
$\{\nu\}=\nu-[\nu]$ and $\bar\theta=\theta-2\pi n$
if $2\pi n\le \theta < 2\pi(n+1)\,\, \, (n\in \mathbb{Z})\,\,$.
This definition of
$\bar\theta$
%We note that
ensures the single-valuedness
of
the above wave function \cite{oren}.
Note also that we have not imposed the exchange symmetry property
of the wave function yet.
In order to get the above solution, one begins with the
partial wave analysis, where one takes
$e^{ipr\cos \theta}$ as an initial wave function.
The resulting partial wave solution with the $\psi(0)=0$ boundary condition
can be summed, leading to the above integral form.

%For the boundary condition $\psi(0)=0$, the transformation
%\be
%\psi(r,\,\theta) \rightarrow \psi'(r,\,\theta)= e^{in\theta}\,\,
%\psi(r,\,\theta)
%\ee
%are well-defined at the origin so that it is a legitimate
%gauge transformation. But for a more general boundary
%condition at the origin discussed below, this cannot
%be a gauge transformation since it is not well defined.

Later, we shall show that  the eigenvalues of $\Omega$ range
over $-2\le \nu \le 2$. When
$1\le |\nu|\le 2$, solely the cases of $\nu=\pm 1,\pm 2$ are relevant for
our application below. We shall treat these four cases separately.
The remaining possibility then lies only within the interval
$(-1,1)$. % and let us focus on this case first.
For this range of $\nu$, one may consider the self-adjoint extension
of the $s$-wave (zero orbital angular momentum) part, which is relevant
for a general understanding of the scattering amplitude. The
extension is dictated by
the boundary condition
\be
\Big[\, r^{|\nu|}\,\psi(r)- R^{2|\nu|}\, {d\,r^{|\nu|}\psi(r)\over d \,
r^{2|\nu|} }\,\Big]_{r=0}=0\,,
\ee
where $R$ is a scale-dependent parameter % introduced which is
related to the
RG-scale of  field theory  \cite{amelino}.

In our case, we also require quantum scale invariance,
which follows from the superconformal invariance of the underlying
theory. This selects only two possibilities. One is
the $R=0$ case or equivalently $\psi(0)=0$, which is called the ``repulsive
critical'' boundary condition.
The other scale-invariant boundary condition
is ${d\,r^{|\nu|}\psi(r)\over d \,
r^{2|\nu|} }\,\big|_{r=0}=0$, which we call the
``attractive critical'' boundary condition.
For the former boundary condition, the scattering solution
was already presented in Eq.~(\ref{ssol}). For the latter, the
scattering solution becomes
\be
\psi(r,\theta)&=& e^{ipr\cos \theta-i\, \nu(\bar{\theta}-\pi)}
-\sin \nu \pi\,
e^{-i ([\nu]+1)\theta}\, \int^\infty_{-\infty}
{dt\over \pi} e^{ipr \cosh t}
{e^{-\{\nu\}t}
\over
e^{-i \theta}- e^{-t}}\nn\\
&-&\big[\,
e^{- {|\nu|\,\pi\, i\over 2}}\, J_{|\nu|}(p r)
-e^{ {|\nu|\,\pi\,i\over 2}}\, J_{-|\nu|}(p r)
\,\big]\,.
\label{ssol1}
\ee
For $\nu=\pm 1,\pm 2$, %only $\psi(0)=0$ boundary condition is
the self-adjoint extension is
not allowed, and the scattering solution becomes
almost trivial:
\be
\psi(r,\theta)&=& e^{ipr\cos \theta-i\, \nu({\theta}-\pi)}\,.
\label{ssol2}
\ee

The scattering amplitude is defined by the asymptotic form
\be
\psi \sim e^{ipr \cos\theta}
+ {1\over \sqrt{r}}\, e^{i(pr +{\pi\over 4})}\, f(\theta)\,
\ee
in the large-$r$ limit.
For $|\nu|< 1$, the scattering amplitude can be found as
\be
f_\pm (\theta)= f_{\rm ns} (\theta)
+f_{\rm s\pm} (\theta)\,,
\ee
where the non-s-wave  and the
s-wave contributions,
$f_{\rm ns} (\theta)$ and $f_{\rm s} (\theta)$,
are, respectively,  given by
\bea
f_{\rm ns} (\theta)&=&{-{i\over \sqrt{2\pi p}}}\,
\Big[
\sin \pi \nu\, \cot {\theta\over 2}+ 2 \sin^2 {\pi\nu\over 2}\,
(1-2\pi\,\delta(\theta))
\Big]
\label{nsamp}
\\
f_{{\rm s}\pm} (\theta)&=&{-{i\over \sqrt{2\pi p}}}\,
\big(\,e^{\mp i  \pi |\nu|}-1\,\big)
\label{samp}\,.
\eea
Here, the upper and the lower signs are, respectively,
for the repulsive and the attractive critical boundary
conditions.
%The first and the second terms are corresponding to
% non s-wave contributions  while the term in the
% last parenthesis is the s-wave part.
The term involving the forward delta function  arises  from
the phase-modulated incoming wave
in Eqs.~(\ref{ssol}),  (\ref{ssol1}) and  (\ref{ssol2}).
Its scattering part has the asymptotic form
\bea
&& ( e^{-i\, \nu({\theta}-\pi)}-1)\, e^{ipr\cos \theta}
\rightarrow  ( e^{-i\, \nu({\theta}-\pi)}-1)\,
{1\over \sqrt{2\pi p r}}\, e^{i(pr -{\pi\over 4})}\,
\sum^\infty_{n=-\infty} e^{in\theta}
\nn\\
&&\ \ \ \ \ \ \ =  ( e^{-i\, \nu({\theta}-\pi)}-1)
{e^{i(pr -{\pi\over 4})}\over \sqrt{2\pi p r}}
\, 2\pi\, \delta(\theta)
=( \cos \nu\pi -1)
{ e^{i(pr -{\pi\over 4})}\over \sqrt{2\pi p r}}
\,2\pi\, \delta(\theta)\,,
\eea
where, for the last equality, we average the $\theta=0$ and the
$\theta=2\pi$ contributions. Though we shall not present
any further details, the presence of the delta function
 can also be checked from the unitarity requirement of the scattering
matrix.
%as well as the analysis based on the wave packet.
For the $\nu=\pm 1,\pm 2$ cases, the scattering amplitude
directly obtained from Eq.~(\ref{ssol2}) agrees with the expression
in Eq.~(\ref{samp}) if the corresponding values are substituted.

Let us now discuss the scattering amplitude from the view point
of a field theory perturbative
analysis. We first note that
the strength of the
contact interaction is not arbitrary. The quantum
scale-invariance requires
\be
\Big({m\, c\over 4}\Big)^2= (\pi \,\nu)^2\,,
\label{beta}
\ee
on which the $\beta$-function for the interaction strength
$c$ vanishes. This aspect will be discussed in detail later,
 but we assume that $c$ takes values
that satisfy this condition of criticality.

From the perturbative analysis, one may show that
the non s-wave part does not arise at all from  amplitudes
that contain more than one contact interaction vertex.
Then, using the equivalence between quantum mechanics
and the non-relativistic field theory,
we claim that
the non-s-wave field theory amplitude should be given by
$f_{\rm ns}(\theta)$ in Eq.~(\ref{nsamp}).

On the other hand, the s-wave part receives
contributions
involving an odd number of the contact interaction vertices.
%at the critical quantum scale invariance case.
%At non scale invariant case,
There can be a dependence on even numbers
of $c$, which can be replaced by the same powers
of $\nu$.
% but this will not be relevant to us.
One may also show that
the $s$-wave part involves only
even powers of $\nu$ without
any odd powers of  $\nu$. Furthermore, the perturbative
result obviously requires that $f_{\rm s}$ should be analytic
in $c$ and $\nu$. Hence, using
Eq.~(\ref{beta}) and the equivalence between the quantum mechanics
and the non-relativistic field theory, the field theoretic {\it critical}
s-wave amplitude should be
\be
f_{\rm s}= -{i\over\sqrt{2\pi p}}\,\Big[\,
-i \sin \big(
{mc\over 4}\big)
+ \cos \pi\nu -1
\Big]\,.
\label{sss}
\ee
One finally imposes the exchange symmetry
\be
f^{\rm full}(\theta)=
f(\theta)+(-)^{F_1 F_2}f(\theta+\pi)\,,
\ee
where $F_1$ and $F_2$ are fermion numbers of
incoming particles.

This result can be extended even for general boundary
conditions, including scale-variant values of $c$, which was first
suggested in Ref.~\cite{amelino} based on the two-loop analysis and was proven
to all-loop orders in Ref.~\cite{Kim:1996rz}. A
similar analysis can be carried out for our problem
starting from the Schr\"odinger equation in Eq.~(\ref{sch}).
One may start from the scattering solution of the Schr\"odinger equation
in Eq.~(\ref{sch}), which is given by the solution in Eq.~(\ref{ssol})
with the replacement of $\nu$ by the matrix $\Omega$.
In this solution, one is defining $|\Omega|$ in the basis
where $\Omega$ is diagonal, which is achieved by diagonalizing
${\rm P}-\overline{\rm P}$.
%It has eigenvalues
%$\pm 1$ with projections of spaces by
%${1\over 2}\,({1\mp {\rm P}\overline{\rm P}})$, i.e.
Its eigenspaces  are %with a single eigenvalue
given by the projections
\be
\Pi^g_0={1\over 2}\big(1+{\rm P}\overline{\rm P}\big)\,,\ \ \ \
\Pi^g_\pm={1\over 4}\Big(\,2\pm({\rm P}-\overline{\rm P}
\,\Big)\,\,{1\over 2}
\big(1-{\rm P}\overline{\rm P}\big)
\ee
with eigenvalues
\be
\Omega\,
\, \Pi^g_0=0\,, \ \ \ \
\Omega\,
\, \Pi^g_\pm=\pm\Big({2\over k}\Big)\, \Pi^g_\pm\,.
\ee
%Therefore, $|\Omega|={2\over k} I$.
The
non-s-wave scattering amplitude is simply given by
\be
F_{ns}(\theta)=
{-{i\over \sqrt{2\pi p}}}\,
\Big[
\sin \pi \Omega\, \cot {\theta\over 2}+ 2 \sin^2 {\pi\Omega\over 2}\,
(1-2\pi\,\delta(\theta))
\Big]\,.
\ee
This can be proven by going to the eigenspace where one may use our previous
result for the single component.
Including the exchange symmetries,
one has
\be
F^{\rm full}_{ns}(\theta)= F_{ns}(\theta)+F_{ns}(\theta+\pi)\mathbb{P}\,
{\rm P} \overline{\rm P}\,.
\ee

For the s-wave contribution, let us first note a few things.
We introduce a new gauge interaction matrix
$\widetilde{\Omega}$ as
\be
\widetilde\Omega = {1\over 2 k}\,\, \big(\mathbb{I}-\mathbb{P}\,\big)\,\otimes
(\,
{\rm P}-\overline{\rm P}
\,)\,,
\ee
which includes the effect of the s-wave exchange symmetry.
Namely, the projection
\be
\Pi^{\rm ex}={1\over 2}\big(
1+\mathbb{P}\,{\rm P}\overline{\rm P}
\big)
\ee
defines the projection of in and out s-wave states, and the operator $\Omega$
acting
upon the s-wave projected state becomes $\widetilde\Omega$:
\be
\Omega \,\Pi^{\rm ex} = \widetilde{\Omega}\, \Pi^{\rm ex}=
\Pi^{\rm ex}\,
\widetilde{\Omega}\,.
\ee

As will be shown in detail later on, the quantum scale-invariance
of our Chern-Simons theory
is maintained by the relation
\be
\Big({m\, C\over 4}\Big)^2=\big(\pi\,
\widetilde\Omega
 \big)^2\,,
\label{scales}
\ee
for which one has to consider the full s-wave contribution, which
respects the exchange symmetry.
The supermatrix
relation Eq.~(\ref{scales})
follows from the property of $C$ in Eq.~(\ref{pro}).

For the eigenspace of the supermatrix $C$, we further introduce
projections
\be
\Pi^f_0={1\over 2}\big(1+\mathbb{P}\big)\,,\ \ \ \
\Pi^f_\pm={1\over 2}\Big(\,1\pm{mk\over 4\pi}\big(\mathbb{S}+
\widetilde{\mathbb{S}}+\mathbb{T}
\big)
\,\Big)\,\,{1\over 2}
\big(\,1-\mathbb{P}\,\big)\,.
\ee
Then, $C$ and $\widetilde{\Omega}$ are simultaneously
diagonalized as
\bea
&& C\,\Pi^g_0=
C\,\Pi^f_0=\widetilde{\Omega} \,\Pi^g_0=
\widetilde{\Omega} \,\Pi^f_0=0\,,
 \nn\\
&& C\,\Pi^f_\pm\,\Pi^g_\pm=(\pm)\,(\pm) {8\pi\over mk}\,
\Pi^f_\pm\,\Pi^g_\pm\,,
\nn\\
&& \widetilde{\Omega}\,\Pi^f_\pm\,\Pi^g_\pm=(+)\,(\pm) {2\over k}\,
\Pi^f_\pm\,\Pi^g_\pm\,,
\eea
where the signatures of the first and the second parentheses
are, respectively, those of the first and second projections.
Since, in the eigenbasis, the previous analysis in Eq.~(\ref{sss})
 is valid, the full s-wave
contribution is given by
\be
F^{\rm full}_{\rm s}= -{2i\over\sqrt{2\pi p}}\,\Big[\,
-i \sin \big(
{mC\over 4}\big)
+ \cos \pi\widetilde{\Omega} -1
\Big]\, \Pi^{\rm ex} \,.
\ee
The exact $2\rightarrow 2$ scattering amplitude is then given
by the scattering matrix
\be
F^{\rm full}=F^{\rm full}_{\rm ns}+F^{\rm full}_{\rm s}\,,
\ee
whose component $\langle {\cal A}'_1  {\cal A}'_2\,|\,F^{\rm full}\,|\,
 {\cal A}_1  {\cal A}_2\rangle$ describes the
 ${\cal A}_1  {\cal A}_2\rightarrow
{\cal A}'_1  {\cal A}'_2$ scattering amplitude
in the component notation.

Using this exact construction, one finds that
the $k=1,2$ cases are special. For $k=1$, the scattering amplitude
vanishes completely. For $k=2$, the scattering amplitude
$F^{\rm full}$ involves only forward and
backward delta-function scattering contributions, which are, respectively,
proportional to $\delta(\theta)$ and $\delta(\theta+\pi)$.
These contributions proportional to delta functions
can be absorbed into the definition of  incoming
waves, which leads to null scattering. This implies that they are simply
 artifacts of our scattering formulation.
Therefore, we conclude that 4-point scattering  is completely trivial
for the special cases of $k=1$ and $2$. This observation is also consistent
with the supersymmetry enhancement observed in Ref.~\cite{okab}.

\section{Perturbative amplitude and $\beta$-function}

In this section, we shall present checks of our exact scattering amplitude
by using a direct field theory computation. Basically, the same perturbative
analysis with an arbitrary gauge group, but with a purely bosonic
matter field appears  in Ref.~\cite{oren}, so  we do not need
any new computation of Feynman diagrams. We need to simply include
the statistical factors arising whenever two fermi fields are
exchanged in the computation of Wick contractions.

Adopting the convention in Ref.~\cite{oren},  we shall use the Coulomb
gauge, adding the terms
\be
{\cal L}_{gf}={1\over \xi}{\rm Tr}\,\big(\nabla\cdot {\bf A}\big)^2 +
{1\over \bar\xi}{\rm Tr}\,
\big(\nabla\cdot \bar{\bf A}\big)^2
\ee
to the Lagrangian in Eq.~(\ref{lagrange}), and take the limit,
$\xi\rightarrow 0$ and $\bar{\xi}\rightarrow 0$. The corresponding
ghost terms
\be
{\cal L}_{gh}={\rm Tr}\,\,\eta^\dagger \Big(\nabla^2\, \eta
+i\, [A_i\,,\,
\partial_i\eta\,]\,
\Big)
+{\rm Tr}\,\,{\bar{\eta}}^\dagger \Big(\nabla^2\, \bar\eta
+i\, [\bar{A}_i\,,\,
\partial_i\bar\eta\,]\,
\Big)
\ee
should also be included. Below,
we shall use  the dimensional regularization with spatial
dimension $d=2-2\epsilon$, plus one dimension
% integrals for
corresponding to the time direction.

Following computation in Ref.~\cite{oren}, one may show that
the gluon self-energy contribution
vanishes identically to the one-loop order. In addition,
propagators for the matter fields do not receive
any higher-order corrections. This basically follows from the
fact that, in non-relativistic theories, pair creations are
not possible at all.

The tree-level 4-point amplitude becomes
\be
A_{(0)}= \Big(
{C\over 2}- {2\pi\,i\over m}\, \Omega\, \cot {\theta\over 2}\,
\Big)
+\Big(
{C\over 2}+ {2\pi\,i\over m}\, \Omega\, \tan {\theta\over 2}\,
\Big)\,\mathbb{P}{\rm P}\overline{\rm P}\,,
\ee
which is consistent with our exact scattering amplitude of the
previous section. For the one-loop order, the regularized
amplitude may be evaluated as
\bea
A_{(1)}&=&{m\over 8\pi}\, \Big(
C^2 - {16\pi^2\over m^2}\, {\widetilde\Omega}^2
\Big)\,\Big({1\over\epsilon}+\ln \big({4\pi\mu^2\over p^2}\big)+i\pi-
\gamma\Big)
\nn\\
&+&
{2\pi^2i\over m}\,\Big(\,\delta(\theta)\,\Omega^2
+
\delta(\theta+\pi)\,\Omega^2\,\mathbb{P}{\rm P}\overline{\rm P}
\Big)\,,
\eea
where $\gamma$ is  Euler's constant.
The amplitude is renormalized by adding a counter-term
corresponding to
\be
C=C_{\rm ren}+\delta C\,,
\ee
with
\be
\delta C=
-{m\over 8\pi}\, \Big(
C_{\rm ren}^2 - {16\pi^2\over m^2}\, {\widetilde\Omega}^2
\Big)
\,\Big({1\over\epsilon}+\ln \,4\pi -
\gamma\Big) \,.
\ee
The renormalized amplitude becomes
\be
A^{\rm ren}_{(0)}= \Big(
{C_{\rm ren}\over 2}- {2\pi\,i\over m}\, \Omega\, \cot {\theta\over 2}\,
\Big)
+\Big(
{C_{\rm ren}\over 2}+ {2\pi\,i\over m}\, \Omega\, \tan {\theta\over 2}\,
\Big)\,\mathbb{P}{\rm P}\overline{\rm P}\,
\ee
and
\bea
A^{\rm ren}_{(1)}={m\over 8\pi}\, \Big(
C_{\rm ren}^2 - {16\pi^2\over m^2}\, {\widetilde\Omega}^2
\Big)\,\Big(\,\ln {\mu^2\over p^2}+i\pi\,\Big)
+
{2\pi^2i\over m}\,\Big(\,\delta(\theta)\,\Omega^2
+
\delta(\theta+\pi)\,\Omega^2\,\mathbb{P}{\rm P}\overline{\rm P}
\Big)\,.
\eea
The one-loop  $\beta$ function for the contact interaction matrix becomes
\be
\beta_C\equiv {d C_{\rm ren}\over d\ln \mu}=
{m\over 4\pi}\, \Big(
C_{\rm ren}^2 - {16\pi^2\over m^2}\, {\widetilde\Omega}^2
\Big)\,,
\ee
which becomes critical if
\be
C_{\rm ren}^2 - {16\pi^2\over m^2}\, {\widetilde\Omega}^2=0\,.
\label{cri}
\ee
This is precisely the relation satisfied by our contact interaction
matrix, which is also consistent with the quantum scale-invariance of
the Chern-Simons theory. Upon using the relation in Eq.~(\ref{cri}), one confirms
that the amplitude up to one loop agrees precisely with that of
our exact analysis.
This analysis can be extended to higher orders \cite{amelino,Kim:1996rz},
but we shall not  go to that direction.

\section{Conclusions}

In this note, we consider the non-relativistic superconformal
Chern-Simons theory with fourteen supersymmetries. We show that the theory
is completely specified by the contact and the statistics
interaction matrices,   $C$ and $\Omega$, which respect the  $\mathfrak{psu}(2|2)$ symmetry
of the underlying theory.

Restricted to the two-body sector of the non-relativistic system,
we derive the corresponding
two-body Schr\"odinger equation. From its $2\rightarrow 2$ scattering solution,
we obtain the exact four-point scattering amplitude %which is
valid to all orders
in $1/N$ and $1/k$. We show that
the scattering amplitude becomes completely trivial when $k=1$ and $2$.
We confirm this scattering amplitude  to one-loop order by using a
field theoretic perturbative
computation. Especially, we  verify  that the beta function for the
contact interaction  vanishes to  one-loop order, which is consistent with the
quantum conformal invariance of the theory.  Extending this
check to two-loop order is straightforward.

It would be interesting to extend our analysis to  cases of less supersymmetric
theories, some of which can be obtained from the mass-deformed ABJM theories by using appropriate
non-relativistic limits. Especially, the `PAAP' theory involves only kinematical
supersymmetries, so an explicit check of quantum conformal invariance by computing its beta function of
contact interaction would be quite interesting.

\section*{Acknowledgments}
DB would like to thank Soo-Jong Rey for the initial collaboration  and
extensive discussions.
This work was supported by a 2009 UOS Academic Research Grant.

\appendix
\section{Non-relativistic Chern-Simons theory}

The maximally supersymmetric non-relativistic Chern-Simons system
is described by \cite{Nakayama:2009cz,Lee:2009mm}
%Similarly for $\mu>0$, one gets
%
\bea
{\cal L} &=&   \, {k \over 4\pi}CS(A)-{k \over 4\pi}
CS(\overline{A})\nonumber \\
&+&{\mbox{Tr}} \Big[ \,\,\,
\phi_a^{\dagger} \Big(\,i\,D_0+ {1\over 2m} \vec{D}^2\Big) \phi^a
+
\tilde\phi_\alpha^{\dagger}
\Big(\,i\,D_0+ {1\over 2m} \vec{D}^2\Big) \tilde{\phi}^\alpha\,\,\, \Big]
\nonumber\\
&+&{\mbox{Tr}}
\Big[\psi_\alpha^{\dagger} \Big(\,i\,D_0+ {1\over 2m} \vec{D}^2\,\Big)
\psi^\alpha
+\tilde{\psi}_a^{\dagger} \Big(\,i\,D_0+ {1\over 2m} \vec{D}^2\,\Big)
\tilde{\psi}^a\Big] -V\,\,,
%\nonumber \\
\eea
where
\be
CS(A)= \epsilon^{\mu\nu\lambda}
\mbox{Tr} \Big(A_\mu \partial_\nu A_\lambda +
{2 i \over 3} A_\mu A_\nu A_\lambda \Big)\,,
\ee
and
\bea
V&=&
{1\over 2m}
{\mbox{Tr}}
\Big[ \,
  (\psi^\alpha \psi^{\dagger}_\alpha - \tilde{\psi}^a
\tilde{\psi}_a^{\dagger})\,
F_{12}(A)
+  ( \psi^{\dagger}_{\alpha} \psi^\alpha  -
\tilde{\psi}^{\dagger}_a
\tilde{\psi}^a )\,
 F_{12}(\overline{A}) \,
\Big]
\nonumber \\
&+&
 {\pi\over mk}
{\mbox{Tr}} \Big[ \,\,
   \phi^{\dagger}_a \phi^a \, \phi^{\dagger}_b \phi^b -
 \phi^a \phi^{\dagger}_a\,\phi^b \phi^{\dagger}_b
-\tilde{\phi}^{\dagger}_\alpha
\tilde{\phi}^\alpha\, \tilde{\phi}^{\dagger}_{\beta}
\tilde{\phi}^\beta +
 \tilde{\phi}^\alpha \tilde{\phi}^{\dagger}_\alpha\,\tilde{\phi}^\beta
\tilde{\phi}^{\dagger}_\beta\,\,
\Big]
\nonumber \\
&+&
 {\pi\over mk}
{\mbox{Tr}} \Big[
  \big( \phi^{\dagger}_a\phi^a-  \tilde{\phi}^{\dagger}_{\alpha}
\tilde{\phi}^\alpha\big)
\big(\psi^{\dagger}_{\beta}\psi^\beta + \tilde{\psi}^{\dagger}_b \tilde\psi^b
\big)
+
 \big( \pa \pad -\ta \tad\big)
\big(\sb \sbd + \ub \ubd\big)
\Big]
\nonumber \\
 &-&
 {2\pi\over mk}
{\mbox{Tr}} \Big[
  \pad \ua \ubd\pb
+ \pa \uad \ub \pbd
+\sad \ta \tbd \sb
+ \sa \tad \tb \sbd
\Big]
\nonumber\\
&-&
 {2\pi\over mk}
{\mbox{Tr}} \Big[
\sad\ta \ubd \pb
+\sa\tad\ub\pbd
+\pad\ua\tbd\sb
+\pa\uad\tb\sbd
\Big]\,.
\eea
The indices
%The SU(2) indices
$a,b$ and $\alpha,\beta$
 are, respectively, for the first and the second $\rm SU(2)$
of $\mathfrak{psu}(2|2)$. All of them  run over $\{1,2\}$. The indices $\mu,\nu,
\lambda$ are for
the spacetime
directions, which run over $\{0,1,2\}$, and %$i,j,k$
we use vector notation
for the spatial
directions.
All the fields are $N\times N$ matrix valued, and the covariant derivative
is defined by
\be
D_\mu \Phi = \partial_\mu \Phi + i A_\mu \Phi - i \Phi \overline{A}_\mu \,
%\, , \quad D_m Y^\dagger_I =
\ee
such that the system possesses the
${\rm U(N)}\times {\rm U}
{\rm (N)}$ gauge symmetry.
We use the Gauss law constraints
\bea
&&{k\over 2\pi} \, F_{12}(A) = \rho_A\equiv \pa \pad+ \ta\tad-\sa\sad-\ua\uad\,,
\nonumber\\
&&{k\over 2\pi} \, F_{12}(\overline{A}) =
\rho_{\bar{A}}\equiv  \pad\pa+ \tad\ta+\sad\sa+\uad\ua\,
\eea
to eliminate the $ F_{12}(A)$ and  the $F_{12}(\overline{A})$ terms.
The potential then becomes
\bea
V=V_{BB}+V_{FF}+V_{BF}+ \widetilde{V}_{BF}\,,
\label{apotential}
\eea
with
\bea
V_{BB}&=& {\pi\over mk}
{\mbox{Tr}} \Big[\,
\pad\pa\pbd \pb - \pa\pad\pb \pbd-
\tad\ta\tbd \tb + \ta\tad\tb \tbd
\,\Big]\,,\nonumber\\
V_{FF}&=& {\pi\over mk}
{\mbox{Tr}} \Big[\,
\sad\sa\sbd \sb - \sa\sad\sb \sbd-
\uad\ua\ubd \ub + \ua\uad\ub \ubd
\,\Big]\,,\nonumber\\
V_{BF}&=&{2\pi\over mk}
{\mbox{Tr}} \Big[\,
\pad\pa\sbd \sb + \pa\pad\sb \sbd-
\tad\ta\ubd \ub - \ta\tad\ub \ubd
\,\Big]\,,\nonumber\\
\widetilde{V}_{BF}&=& -{2\pi\over mk}
{\mbox{Tr}} \Big[\,
\pad\ua\ubd \pb + \pa\uad\ub \pbd+
\sad\ta\tbd \sb + \sa\tad\tb \sbd\nonumber\\
&&\ \ \ \ \ \ \ \ \ \ \ +
\sad\ta\ubd \pb + \sa\tad\ub \pbd+
\pad\ua\tbd \sb + \pa\uad\tb \sbd
\,\Big]\,.
\eea
This is the form of the  potential we use in the main body of this paper.

\end{document}